\documentclass[aps, preprintnumbers, nofootinbibt, onecolumn, preprint]{revtex4-1}
\usepackage{amsmath}
\usepackage{eurosym}
\usepackage{amssymb}
\usepackage{graphicx}
\usepackage{color}

\def\be{\begin{equation}}
\def\ee{\end{equation}}
\def\bea{\begin{eqnarray}}
\def\eea{\end{eqnarray}}

\begin{document}

\title{Computation of the general relativistic perihelion precession and of
light deflection via the Laplace-Adomian Decomposition Method}
\author{Man Kwong Mak}
\email{mankwongmak@gmail.com}
\affiliation{Departamento de F\'{\i}sica, Facultad de Ciencias Naturales, Universidad de
Atacama, Copayapu 485, Copiap\'o, Chile}
\author{Chun Sing Leung}
\email{chun-sing-hkpu.leung@polyu.edu.hk}
\affiliation{Department of Applied Mathematics, Polytechnic University of Hong Kong, Hong
Kong SAR}
\author{Tiberiu Harko}
\email{t.harko@ucl.ac.uk}

\affiliation{Department of Physics, Babes-Bolyai University, Kogalniceanu Street,
Cluj-Napoca 400084, Romania,}
\affiliation{School of Physics, Sun Yat-Sen University, Guangzhou 510275, People's
Republic of China}
\affiliation{Department of Mathematics, University College London, Gower Street, London
WC1E 6BT, United Kingdom}

\date{\today }

\begin{abstract}
We study the equations of motion of the massive and massless particles in
the Schwarzschild geometry of general relativity by using the
Laplace-Adomian Decomposition Method, which proved to be extremely
successful in obtaining series solutions to a wide range of strongly
nonlinear differential and integral equations. After introducing a general
formalism for the derivation of the equations of motion in arbitrary
spherically symmetric static geometries, and of the general mathematical
formalism of the Laplace-Adomian Decomposition Method, we obtain the series
solution of the geodesics equation in the Schwarzschild geometry. The
truncated series solution, containing only five terms, can reproduce the
exact numerical solution with a high precision. In the first order of
approximation we reobtain the standard expression for the perihelion
precession. We study in detail the bending angle of light by compact objects
in several orders of approximation. The extension of this approach to more
general geometries than the Schwarzschild one is also briefly discussed.
\end{abstract}

\maketitle
\tableofcontents

\section{Introduction}

General relativity is a very successful theory of the gravitational field,
whose predictions are in excellent agreement with a large number of
astronomical observations and experiments performed at the scale of the
Solar System. In particular, three fundamental tests of general relativity,
the perihelion precession of planet Mercury \cite{x0,x1}, the bending of
light by the Sun \cite{x2,x21}, and the radar echo delay experiment \cite%
{x3,x31} have all fully confirmed, within the range of
observational/experimental errors, the predictions of Einstein's theory of
gravity. But the importance of these effects goes far beyond the limits of
the Solar System. A fast full general relativistic method to simultaneously
constrain the mass of massive black holes, their spin, and the spin
direction by considering both the motion of a star and the propagation of
photons from the star to a distant observer was developed in \cite{x4}. The
spin-induced effects on the projected trajectory and redshift curve of a
star depend on both the value and the direction of the spin. The maximum
effects over a full orbit can differ by a factor up to more than one order
of magnitude for cases with significantly different spin directions. In \cite%
{x5} it was shown that the spin of the massive black hole at the Galactic
Center can be constrained with 1$\sigma$ error $<\sim 0.1$ or even greater
than 0.02 by monitoring the orbital motion of a star with semi major axis $%
<\sim 300$ AU and eccentricity $>\sim 0.95$ over a period shorter than a
decade through future facilities. An improvement in astrometric precision
would be more effective at improving the quality of constraining the spin
than an improvement in velocity precision. Short-period stars orbiting
around the supermassive black hole in our Galactic Center can successfully
be used to probe the gravitational theory in a strong regime \cite{x6}. By
using 19 years of observations of the two best measured short-period stars
orbiting our Galactic Center constraints on a hypothetical fifth force that
arises in some extended theories of gravity or in some models of dark matter
and dark energy were obtained in \cite{x6}. No deviations from General
Relativity were found, and the fifth force strength wss restricted to an
upper 95\% confidence limit of $|\alpha|<0.016$ at a length scale of $%
\lambda = 150$ astronomical units. Moreover, 95\% confidence upper limit on
a linear drift of the argument of periastron of the short-period star S0-2
was obtained, a result that opens the possibility of testing gravitational
theories using orbital dynamic in the strong gravitational regime of a
supermassive black hole. The S-star cluster in the Galactic center allows
the study of physics close to a supermassive black hole, including
distinctive dynamical tests of general relativity \cite{x7}, where a new and
practical method for the investigation of the relativistic orbits of stars
in the gravitational field near Sgr A* was developed, by using a first-order
post- Newtonian approximation to calculate the stellar orbits with a broad
range of periapse distance $r_p$. For in depth discussions of the
experimental and Solar System tests of general relativity see \cite{x8} and
\cite{x81}, respectively.

Due to its importance in many applications, the study of the motion of
massive or massless particles in different geometries obtained as solutions
of Einstein's gravitational field equations, and of their extensions, is a
fundamental field of general relativity. The first exact solution of the
vacuum field equations was the Schwarzschild solution \cite{LaLi}, which can
be used efficiently to explain all astronomical observations at the scale of
the Solar System. The exact equation of motion in Schwarzschild geometry is
highly nonlinear, and therefore to obtain the observable physical parameters
approximate methods must be used. The first order approximation of the
equation of motion already gives the correct approximation of the perihelion
precession of Mercury, and of the deflection of light by the Sun \cite{x8}.
However, due to the importance of the problem many mathematical techniques
for the study of the astrometric properties of the planetary motions and of
the light have been developed. A standard approach is based on the solution
of the Hamilton-Jacobi equation \cite{LaLi},
\begin{equation}
g^{\mu \nu }\frac{\partial S}{\partial x^{\mu }}\frac{\partial S}{\partial
x^{\nu }}-m^{2}c^{2}=0,
\end{equation}%
where $g^{\mu \nu }$ are the components of the metric tensor, and $m$ is the
mass of the particle. By representing $S$ as $S=-Et+M\varphi +S_{r}(r)$,
where $E$ and $M$ are the constants of the energy and angular momentum, one
can obtain the full solution of the equation of motion in Schwarzschild
geometry in an integral form as \cite{LaLi}
\begin{equation}
ct=\frac{E}{mc^{2}}\int {\frac{dr}{\left( 1-\frac{r_{g}}{r}\right) \left[
\left( \frac{E}{mc^{2}}\right) ^{2}-\left( 1+\frac{M^{2}}{m^{2}c^{2}r^{2}}%
\right) \left( 1-\frac{r_{g}}{r}\right) \right] ^{1/2}}},
\end{equation}

\begin{equation}
\varphi =\int \frac{Mdr}{r^{2}\sqrt{\frac{E^{2}}{c^{2}}-\left( m^{2}c^{2}+%
\frac{M^{2}}{r^{2}}\right) \left( 1-\frac{r_{g}}{r}\right) }}.
\end{equation}

The parameters of the orbits can be obtained by solving the integrals by
using some approximate methods. The geodesic equations obtained from the
Schwarzschild gravitational metric in the presence of a cosmological
constant were solved exactly, and expressed in a closed form in \cite{Kran}
as
\begin{equation}
u=\frac{1}{r}=\frac{4}{\alpha_S}\wp(\varphi+\epsilon)+\frac{1}{3 \alpha_S},
\end{equation}
where $\alpha _S=2GM/c^2$, $\epsilon$ is an arbitrary integration constant, $%
\wp$ is the Weierstrass function that gives the inversion of the elliptic
integral $\int_{}^U \frac{dU}{\sqrt{4 U^3-g_2 U-g_3}}=\phi$, by the
Weierstrass function, $U=\wp(\phi+\epsilon)$. In this approach the exact
expression of the perihelion precession is given by $\Delta
_{\omega}=2\left(\omega -\pi\right)$, where $\omega =\int_{e_1}^{\infty}{%
\frac{dt}{\sqrt{4t^3-g_2t-g3}}}$, and $e_1$ is a root of the cubic equation
in the integral. The perihelion precession and deflection of light have been
investigated in the four-dimensional general spherically symmetric spacetime
in \cite{ppc}, where the master equation has also been obtained. As an
application of this master equation, the Reissner-Nordstorm solution and
Clifton-Barrow solution in $f(R)$ gravity have been investigated. The
homotopy perturbation method, which was introduced in \cite{Hom}, was
applied for calculating the perihelion precession angle of planetary orbits
in General Relativity in \cite{Schi1,Schi2}. The basic ideas behind the
homotopy perturbation method are as follows \cite{Hom}. We start from the
nonlinear differential equation $A(u)=g(r), \,\,\, r \in \mathit{\Omega}$,
where $A$ is a general differential operator, and $g(r)$ is a known analytic
function, with the boundary conditions $B(u, \partial u/\partial n)= 0;\, r
\in \mathit{\Gamma}$, where $B$ is a boundary operator, and $\mathit{\Gamma}$
is the boundary of the domain $\mathit{\Omega}$. We assume that the operator
$A$ can be divided into two parts $M$ and $N$, and we reformulate our
initial equation as $M(u) + N(u)= g(r)$. Then the homotopy $v(r, p): \mathit{%
\Omega}\times [0,1] \to \mathit{I\!\!R} $ is constructed in the following
way: $H(v, p)=(1- p) [M(v)- M(y_0)]+p\,[A(v)- g(r)] = 0$, where $r \in
\mathit{\Omega}$ and $p \in [0, 1]$ is an imbedding parameter, and $y_0$ is
an initial approximation of the equation. Since $0 \leq p \leq 1$, it can be
considered as a small parameter, and one can assume that the solution of the
equation can be expressed as a power series in $p$ as $v=\sum_{i=0}^{\infty}{%
v_ip^i}$ When $p \to 1$, then this series becomes the approximate solution
of the equation, that is $u(x)= \lim_{p \to 1} v =\sum_{i=0}^{\infty}{v_i}$.
This series is generally convergent.

The study and the applications of Adomian's Decomposition Method (ADM) \cite%
{R1,R2,b2,b3}, which allows to investigate the solutions of many kinds of
ordinary, partial, stochastic differential and integral equations that
describe numerous physical and/or mathematical problems has attracted a lot
of attention in recent years. Historically, the ADM was first proposed and
applied in the 1980's \cite{new1,new2,new3,new4}. An essential advantage of
the ADM is that with its use one can obtain analytical approximations to the
solutions of a rather wide class of nonlinear (and stochastic) differential
and integral equations, without the need of linearization, perturbation,
closure approximations, or discretization methods. Usually the application
of these methods could lead to the necessity of intensive numerical
computation. Moreover, to make solvable and to obtain closed-form analytical
solutions of a nonlinear problem implies the necessity of introducing some
simplifying and restrictive assumptions.

It is important to mention that ADM can generate the solution of a given
equation in the form of a power series. The terms of the series are obtained
by recursive relations using the Adomian polynomials. Another important
advantage of the ADM is that usually the series solution of the differential
equation converges fast, and therefore the use of this method saves a lot of
computational time. Moreover, in the ADM there is no need to linearize or
discretize the differential equation. Reviews of ADM and its applications in
applied mathematics and physics can be found in \cite{R1,R2}, respectively.
Many studies have been devoted to the modification and improvement of the
ADM in an attempt to increase its accuracy, and/or to extend the
applicability of the initial method \cite%
{b2,b3,b5,b6,b7a,b8,b9,b10,b11,b12,b13,b14,b28,b30,b31,p1,p2}. An important
improvement of the ADM s represented by the Laplace-Adomian Decomposition
method \cite{p3}, in which the Adomian Decomposition Method is applied not
to initial equation, but to the Laplace transformed one.

Even that the ADM has been extensively used in the study of many problems in
different fields of physics and engineering, it has been applied very little
in astronomy, astrophysics or general relativity, the only exception from
this "rule" known to authors being the papers \cite{Ai} and \cite{Ai1}. It
is the purpose of the present paper to investigate the equations describing
the perihelion precession and light bending in general relativity for static
gravitational fields by using the Adomian Decomposition Method, representing
a very powerful mathematical method for the investigation of the solutions
of nonlinear differential equations. For the geometry outside a compact,
stellar type object (the Sun) we adopt some specific static and spherically
symmetric vacuum solutions of general relativity. In particular the power
series solution of the equation describing the motion of massive and
massless particles in Schwarzschild geometry is investigated in detail. As a
first step in our analysis we derive the equations of motion for particles
in arbitrary spherically symmetric spacetimes, and we develop a general
formalism for obtaining the equations of motion that can be used for any
given metric. As the next step in our study we adopt the Schwarzschild form
of the metric, and we apply the Laplace-Adomian decomposition method to
obtain its approximate analytical power series solution for both massive and
massless particles. We compare our solutions with the exact numerical
solutions of the equations of motion, and it turns out that by truncating
our series to five terms only we obtain a very good description of the
solution of the equation of motion. Moreover, in the first approximation we
can reobtain easily the standard expressions of the perihelion precession
and the bending angle of light.

The present paper is organized as follows. We derive the equations of motion
of massive and massless particles in arbitrary static spherically symmetric
spacetimes in Section~\ref{sect1}. The application of the Laplace-Adomian
Decomposition Method to the case of second order nonlinear differential
equations is presented in Section~\ref{sect2}. The power series solution of
the equation of motion of massive particles by using the Laplace-Adomian
Decomposition Method is obtained in Section~\ref{sect3}, where the
comparison with the exact numerical solution is also performed. The motion
of photons in Schwarzschild geometry is investigated in Section~\ref{sect4}.
Finally, in Section~\ref{sect5}, we discuss and conclude our results.

\section{Particle motion in arbitrary spherically symmetric static
space-times}

\label{sect1}

In the following we will restrict our analysis to the case of static and
spherically symmetric metrics, given by
\begin{equation}
ds^{2}=-e^{\nu (r)}dt^{2}+e^{\lambda (r)}dr^{2}+r^{2}d\Omega ^{2},
\label{metr1}
\end{equation}
where we have denoted $d\Omega ^{2}=d\theta ^{2}+\sin ^{2}\theta d\varphi
^{2}$, and $\theta $ and $\varphi $ are the standard coordinates on the
three-sphere. The time variable takes real values only, $t\in R$ while the
radial coordinate $r$ ranges over a finite open interval $r\in \left(
r_{\min },r_{\max }\right) $, so that $-\infty \leq r_{\min }\leq r_{\max
}\leq \infty $. Moreover, we also require that the functions $\nu (r)$ and $%
\lambda (r)$ are strictly positive, and that on the interval $\left( r_{\min
},r_{\max }\right) $ they are (at least piecewise) differentiable. This form
of the metric is relevant for the study of the dynamics of particles (both
massive and massless) in the Solar System.

Important observational evidence for the correctness of the theory of
general relativity is provided, at the level of the Solar System, by three
fundamental tests, which also allow the testing of its extensions and
generalization, as well as of alternative theories of gravitation. These
three essential tests are the perihelion precession of Mercury, the
deflection of photons by the Sun, and the radar echo delay observations.
These three effects have been successfully used to test the Schwarzschild
solution of general relativity, as well as other predictions of the theory.
However, it is also important to study these physical phenomena in arbitrary
static spherically symmetric space-times for any given metric. In the
present Section, we develop a formalism that can be used for obtaining the
equations of motion, and compute the perihelion precession and light bending
angle in any static spherically symmetric metric. This formalism was first
introduced to study the Solar System tests for some modified gravity vacuum
solutions in \cite{Ha1,Ha2,Ha3,Ha4}.

\subsection{The equation of motion of massive test particles}

The geodesic equations of motion of a massive test particle in the
gravitational field of the metric given by Eq. (\ref{metr1}) can be derived
with the use of the variational principle
\begin{equation}
\delta \int \sqrt{e^{\nu }c^{2}\dot{t}^{2}-e^{\lambda }\dot{r}
^{2}-r^{2}\left( \dot{\theta}^{2}+\sin ^{2}\theta \dot{\varphi}^{2}\right) }
ds=0,  \label{var}
\end{equation}
where by a dot we have denoted $d/ds$. It can be easily checked that the
orbit is planar, and therefore without any loss of generality we can take $%
\theta =\pi /2$. Hence $\varphi $ is the only the angular coordinate in this
problem. Since $t$ and $\varphi $ do not appear explicitly in Eq. (\ref{var}%
), their conjugate momenta gives two constants of motion, denoted $E$ and $L$%
, so that
\begin{equation}
e^{\nu }c^{2}\dot{t}=E=\mathrm{constant}, \qquad r^{2}\dot{\varphi}=L=%
\mathrm{constant}.  \label{consts}
\end{equation}
The constant $E$ gives the energy of the particle, while the constant $L$ is
related to its angular momentum.

From the line element Eq. (\ref{metr1}) we obtain the following equation of
motion for $r$,
\begin{equation}
\dot{r}^{2}+e^{-\lambda }r^{2}\dot{\varphi}^{2}=e^{-\lambda }\left( e^{\nu
}c^{2}\dot{t}^{2}-1\right) \,.  \label{constants}
\end{equation}

By substituting $\dot{t}$ and $\dot{\varphi}$ from Eqs. (\ref{consts}) gives
the following relation,
\begin{equation}
\dot{r}^{2}+e^{-\lambda }\frac{L^{2}}{r^{2}}=e^{-\lambda }\left(\frac{E^{2}
}{c^{2}}e^{-\nu }-1\right) .  \label{inter1}
\end{equation}

We introduce now a new variable $u$, defined as $r=1/u$, as well as the
transformation $d/ds=Lu^{2}d/d\varphi $. Then Eq. (\ref{inter1}) takes the
form
\begin{equation}
\left( \frac{du}{d\varphi }\right) ^{2}+e^{-\lambda }u^{2}=\frac{1}{L^{2}}
e^{-\lambda }\left( \frac{E^{2}}{c^{2}}e^{-\nu }-1\right) .
\end{equation}

We represent formally $e^{-\lambda }$ as $e^{-\lambda }=1-f(u)$, thus
obtaining
\begin{equation}
\left( \frac{du}{d\varphi }\right) ^{2}+u^{2}=f(u)u^{2}+\frac{E^{2}}{%
c^{2}L^{2}} e^{-\nu -\lambda }-\frac{1}{L^{2}}e^{-\lambda }\equiv G(u).
\label{ueq_basic}
\end{equation}

We take the derivative of the above equation with respect to $\varphi $,
which gives
\begin{equation}
\frac{d^{2}u}{d\varphi ^{2}}+u=F(u),  \label{inter2}
\end{equation}
where
\begin{equation}
F(u)=\frac{1}{2}\frac{dG(u)}{du}.
\end{equation}

Eq.~(\ref{inter2}) gives the equation of motion of a particle in an
arbitrary spherically symmetric geometry.

\subsubsection{The precession of the perihelion}

The root of the equation $u_{0}=F\left( u_{0}\right) $ gives a circular
orbit with $u=u_{0}$. Any deviation $\delta =u-u_{0}$ from it can be
obtained from the substitution of $u=u_{0}+\delta $ into Eq. (\ref{inter2}),
which gives the equation
\begin{equation}
\frac{d^2\delta }{d\varphi ^2}+u_0+\delta =F\left(u_0+\delta\right).
\end{equation}
In the first order of approximation $F\left(u_0+\delta\right)\approx
F\left(u_0\right)+\left.\left(dF/du\right)\right|_{u=u_0}\delta+O\left(
\delta ^{2}\right)$, and hence
\begin{equation}
\frac{d^{2}\delta }{d\varphi ^{2}}+\left[ 1-\left( \frac{dF}{du}\right)
_{u=u_{0}}\right] \delta =O\left( \delta ^{2}\right).
\end{equation}

Therefore, to first order in $\delta $, the trajectory of the massive
particle can be obtained as
\begin{equation}
\delta =\delta _{0}\cos \left( \sqrt{1-\left( \frac{dF}{du}\right) _{u=u_{0}}%
}\varphi +\beta \right) ,
\end{equation}%
where $\delta _{0}$ and $\beta $ are two arbitrary constants of integration.
The angles for which $r$ is minimum are the angles of the perihelia of the
orbit. Therefore they are determined from the condition that $u$ or $\delta $
is maximum. Hence, from one perihelion to the next the orbital angle varies
by a quantity $\Delta \varphi $, given by
\begin{equation}
\Delta \varphi =\frac{2\pi }{\sqrt{1-\left( \frac{dF}{du}\right) _{u=u_{0}}}}%
=\frac{2\pi }{1-\sigma }.  \label{prec}
\end{equation}

The parameter $\sigma $ introduced in the previous equation is called the
perihelion advance. From a physical point of view it represents the rate of
advance of the perihelion after one rotation. As the test particle advances
through $\varphi $ radians in its orbit, its perihelion precesses by $\sigma
\Delta \varphi $ radians. From Eq. (\ref{prec}), $\sigma $ can be expressed
as
\begin{equation}
\sigma =1-\sqrt{1-\left( \frac{dF}{du}\right) _{u=u_{0}}},
\end{equation}
or, for small $\left( dF/du\right) _{u=u_{0}}$, as
\begin{equation}
\sigma =\frac{1}{2}\left( \frac{dF}{du}\right) _{u=u_{0}}.
\end{equation}

For a complete rotation of the planet we obtain $\varphi \approx 2\pi
(1+\sigma )$. Hence the advance of the perihelion is
\begin{equation}
\delta \varphi =\varphi -2\pi \approx 2\pi \sigma .
\end{equation}
To be able to obtain effective estimations of the perihelion precession we
must know the expression of the angular momentum of the particle $L$ as a
function of the geometric parameters of the orbit. We will obtain now the
expression of $L$ in the Newtonian limit \cite{Ha1,Ha2,Ha3,Ha4}. 

Let's assume that the planet moves
on a Keplerian ellipse, with semi-axis $\bar{a}$ and $\bar{b}$,
respectively, where $\bar{b}=\bar{a}\sqrt{1-e^{2}}$, and by $e$ we have
denoted the eccentricity of the orbit. The ellipse has a surface area $\pi
\bar{a}$ $\bar{b}$. The oriented surface area of the ellipse is $d\vec{A}%
=\left( \vec{r}\times d\vec{r}\right) /2 $, and consequently the areolar
velocity of the planet is given by $\left\vert d\vec{A}/dt\right\vert
=\left\vert \vec{r}\times d\vec{r}\right\vert /2=r^{2}\left( d\varphi
/dt\right) /2\approx \pi \bar{a}^{2}\sqrt{1-e^{2}}/T$ , where $T$ is the
period of the planetary motion. On the other hand $T$ can be obtained from
Kepler's third law as $T^{2}=4\pi ^{2}\bar{a}^{3}/GM$ \cite{LaLim}. In the
Newtonian limit of small velocities $ds\approx cdt$, and the conservation
equation of the angular momentum reduces to $r^{2}d\varphi /dt=cL$. Hence we
obtain first $L=2\pi \bar{a}^{2}\sqrt{1-e^{2}}/cT$, and hence
\begin{equation}
\frac{1}{L^{2}}=\frac{c^{2}}{GM\bar{a}\left( 1-e^{2}\right) }.  \label{L2}
\end{equation}

\subsubsection{The equation of motion of massive particles in Schwarzschild
geometry}

As a first astronomical application of the formalism introduced in the
previous Section we obtain the precession of the perihelion of a planet in
the Schwarzschild geometry, with
\begin{equation}
e^{\nu }=e^{-\lambda }=1-2GM/c^{2}r=1-\left( 2GM/c^{2}\right) u.
\end{equation}
Then we immediately obtain $f(u)=\left( 2GM/c^{2}\right) u$. On the other
hand since for this geometry $\nu +\lambda =0$, we easily find
\begin{equation}
G(u)=\frac{2GM}{c^{2}}u^{3}+\frac{1}{L^{2}}\left( \frac{E^{2}}{c^{2}}%
-1\right) +\frac{2GM}{c^{2}L^{2}}u,
\end{equation}%
and
\begin{equation}
F(u)=3\frac{GM}{c^{2}}u^{2}+\frac{GM}{c^{2}L^{2}},
\end{equation}
respectively. Therefore the equation of motion of a massive test particle in
Schwarzschild geometry is given by
\begin{equation}
\frac{d^{2}u}{d\varphi ^{2}}+u=3\frac{GM}{c^{2}}u^{2}+\frac{GM}{c^{2}L^{2}}.
\label{eqSc}
\end{equation}

The radius $u=u_0$ of the circular orbit is found as the solution of the
algebraic equation
\begin{equation}
u_{0}=3\frac{GM}{c^{2}}u_{0}^{2}+\frac{GM}{c^{2}L^{2}},
\end{equation}
with the only physically acceptable solution given by
\begin{equation}
u_{0}=\frac{1\pm \sqrt{1-12G^{2}M^{2}/c^{4}L^{2}}}{6GM/c^{2}}\approx \frac{%
GM }{c^{2}L^{2}}.
\end{equation}

Therefore
\begin{equation}
\delta \varphi =\pi \left( \frac{dF}{du}\right) _{u=u_{0}}=\frac{6\pi GM}{%
c^{2}\bar{a}\left( 1-e^{2}\right) },
\end{equation}%
which is the standard general relativistic result \cite{LaLi}.

\subsection{Equation of motions of photons and the deflection of light}

In a gravitational field a photon follows a null geodesic, given by $%
ds^{2}=0 $. In this case the affine parameter along the trajectory of the
photon can be taken as an arbitrary quantity $\tilde{\lambda}$. In the
following we denote again by a dot the derivatives with respect to it.
Similarly to the case of the motion of massive particles we have two
constants of motion, the energy $E$ and the angular momentum $L$, which can
be obtained from Eqs. (\ref{consts}).

The equation of motion of the photon is given by
\begin{equation}
\dot{r}^{2}+e^{-\lambda }r^{2}\dot{\varphi}^{2}=e^{\nu -\lambda }c^{2}\dot{t}
^{2}.
\end{equation}
By using the constants of motion the above equation can be transformed into
\begin{equation}
\dot{r}^{2}+e^{-\lambda}\frac{L^{2}}{r^{2}}=\frac{E^{2}}{c^{2}}e^{-\nu
-\lambda }.
\end{equation}

We change now the independent variable $r$ to $u=1/r$. With the use of the
conservation equations we eliminate the derivative with respect to the
affine parameter, thus obtaining
\begin{equation}
\left( \frac{du}{d\varphi }\right) ^{2}+u^{2}=f(u)u^{2}+\frac{1}{c^{2}}\frac{%
E^{2}}{L^{2}}e^{-\nu -\lambda }\equiv P(u)\,.  \label{P_eq_basic}
\end{equation}%
We take the derivative of Eq.~(\ref{P_eq_basic}) with respect to $\varphi $,
and thus we find the basic equation of the photon in an arbitrary static
spherically symmetric geometry, as given by
\begin{equation}
\frac{d^{2}u}{d\varphi ^{2}}+u=V(u),  \label{phot}
\end{equation}%
where we have denoted
\begin{equation}
V(u)=\frac{1}{2}\frac{dP(u)}{du}.
\end{equation}

In the particular case of the Schwarzschild geometry we have $\nu +\lambda
=0 $ and $f(u)=\left( 2GM/c^{2}\right) u$, giving $P(u)=\left(
2GM/c^{2}\right) u^{3}$ and $V(u)=\left( 3GM/c^{2}\right) u^{2}$,
respectively. Hence the equation of motion of photons in the Schwarzschild
metric is obtained as
\begin{equation}
\frac{d^{2}u}{d\varphi ^{2}}+u=\frac{3GM}{c^{2}} u^{2}.
\end{equation}

\subsubsection{The deflection angle of light}

In the lowest order of approximation we can neglect the term on the right
hand side of Eq.~(\ref{phot}). Then the solution is given by a straight
line,
\begin{equation}  \label{u0}
u=\frac{\cos \varphi }{R},
\end{equation}
where by $R$ we have denoted the distance of the closest approach to the
central massive gravitating object. In the next order of approximation Eq. (%
\ref{u0}) is substituted into the right-hand side of Eq. (\ref{phot}). Hence
the equation of the trajectory is given by a second order linear
inhomogeneous differential equation,
\begin{equation}  \label{uQ}
\frac{d^{2}u}{d\varphi ^{2}}+u=V\left( \frac{\cos \varphi }{R}\right),
\end{equation}
which has a general solution given by $u=u\left( \varphi \right) $. The
photons travel towards the star from infinity at the asymptotic angle $%
\varphi =-\left( \pi /2+\varepsilon \right) $, and are deflected to infinity
at the asymptotic angle $\varphi =\pi /2+\varepsilon $. The angle $%
\varepsilon $ can be computed by solving the algebraic equation $u\left( \pi
/2+\varepsilon \right) =0$. For the total deflection angle of the photon
beam we find $\delta =2\varepsilon $.

\paragraph{The deflection of light in Schwarzschild geometry.}

We consider now the case of the Schwarzschild geometry. In the lowest order
of approximation from Eqs. (\ref{u0}) and (\ref{uQ}) we obtain for the
photon trajectory the second order linear differential equation
\begin{equation}
\frac{d^{2}u}{d\varphi ^{2}}+u=\frac{3GM}{c^{2}R^{2}}\cos ^{2}\varphi =\frac{%
3GM}{ 2c^{2}R^{2}}\left( 1+\cos 2\varphi \right) ,
\end{equation}
having the general solution given by
\begin{equation}
u=\frac{\cos \varphi }{R}+\frac{3GM}{2c^{2}R^{2}}\left( 1-\frac{1}{3}\cos
2\varphi \right) .  \label{ulight}
\end{equation}

By substituting $\varphi =\pi /2+\varepsilon $, $u=0$ into Eq. (\ref{ulight}%
) we easily find
\begin{equation}
\varepsilon =\frac{2GM}{c^{2}R},
\end{equation}
where we have used the simple trigonometric relations $\cos \left( \pi
/2+\varepsilon \right) = -\sin \varepsilon $, $\cos \left( \pi +2\varepsilon
\right) = -\cos 2\varepsilon $, and the approximations $\sin \varepsilon
\approx \varepsilon $ and $\cos 2\varepsilon \approx 1$, respectively. The
total deflection angle of light in the Schwarzschild geometry is thus $%
\delta =2\varepsilon =4GM/c^{2}R$, a well-known result in general relativity
\cite{LaLi}.

\section{The Laplace-Adomian method for nonlinear second order ordinary
differential equations}

\label{sect2}

\subsection{The general formalism}

Let's consider a nonlinear differential equation of the form
\begin{equation}
y^{\prime \prime }+\omega ^{2}y+b^{2}+g(y)=0,  \label{eqb}
\end{equation}%
where $\omega $ and $b$ are constants, and $g$ is an arbitrary nonlinear
function of dependent variable $y$. Eq.~(\ref{eqb}) must be integrated with
the initial conditions $y(0)=y_{0}=a$, and $y^{\prime }(0)=0$, respectively.

In the Laplace-Adomian method we first apply the Laplace transformation
operator $\mathcal{L}$ to Eq.~(\ref{eqb}), thus obtaining
\begin{equation}
\mathcal{L}\left[ y^{\prime \prime }\right] +\omega ^{2}\mathcal{L}[y]+%
\mathcal{L}[b^{2}]+\mathcal{L}\left[ g(y)\right] =0.
\end{equation}

With the use of the properties of the Laplace transformation we easily find
\begin{equation}
\left( s^{2}+\omega ^{2}\right) \mathcal{L}[y]-sy(0)-y^{\prime }(0)+\frac{%
b^{2}}{s}+\mathcal{L}\left[ g(y)\right] =0.
\end{equation}

With the use of the initial conditions for our problem we obtain
\begin{equation}
\mathcal{L}[y]=\frac{as}{s^{2}+\omega ^{2}}-\frac{b^{2}}{s\left(
s^{2}+\omega ^{2}\right) }-\frac{1}{s^{2}+\omega ^{2}}\;\mathcal{L}[g(y)].
\label{6a}
\end{equation}

As a next step in our analysis we assume that the solution can be
represented in the form of an infinite series,
\begin{equation}
y(x)=\sum_{n=0}^{\infty }y_{n}(x),  \label{7a}
\end{equation}%
where the terms $y_{n}(x)$ are computed recursively. As for the nonlinear
operator $g(y)$, it is decomposed as
\begin{equation}
g(y)=\sum_{n=0}^{\infty }A_{n},  \label{8a}
\end{equation}%
where the $A_{n}$'s are the so-called Adomian polynomials, defined generally
as \cite{R2}
\begin{equation}
A_{n}=\left. \frac{1}{n!}\frac{d^{n}}{d\epsilon ^{n}}f\left(
\sum_{i=0}^{\infty }{\epsilon ^{i}y_{i}}\right) \right\vert _{\epsilon =0}.
\end{equation}

The first five Adomian polynomials can be obtained in the following form,
\begin{equation}
A_{0}=f\left( y_0\right) ,  \label{Ad0}
\end{equation}%
\begin{equation}
A_{1}=y_{1}f^{\prime }\left( y_0\right) ,  \label{Ad1}
\end{equation}%
\begin{equation}
A_{2}=y_{2}f^{\prime }\left( y_0\right) +\frac{1}{2}y_{1}^{2}f^{\prime
\prime }\left( y_0\right) ,  \label{Ad2}
\end{equation}%
\begin{equation}
A_{3}=y_{3}f^{\prime }\left( y_0\right) +y_{1}y_{2}f^{\prime \prime }\left(
y_0\right) +\frac{1}{6}y_{1}^{3}f^{\prime \prime \prime }\left( y_0\right) ,
\label{Ad3}
\end{equation}
\begin{equation}
A_{4}=y_{4}f^{\prime }\left( y_0\right) +\left[ \frac{1}{2!}%
y_{2}^{2}+y_{1}y_{3}\right] f^{\prime \prime }\left( y_0\right) +\frac{1}{2!}%
y_{1}^{2}y_{2}f^{\prime \prime \prime }\left( y_0\right) +\frac{1}{4!}%
y_{1}^{4}f^{(\mathrm{iv})}\left( y_0\right) .  \label{Ad4}
\end{equation}

Substituting Eqs. (\ref{7a}) and (\ref{8a}) into Eq. (\ref{6a}) we obtain
\begin{equation}
\mathcal{L}\left[ \sum_{n=0}^{\infty }y_{n}(x)\right] =\frac{as}{%
s^{2}+\omega ^{2}}-\frac{b^2}{s\left( s^{2}+\omega ^{2}\right) }-\frac{1}{%
s^{2}+\omega ^{2}}\mathcal{L}[\sum_{n=0}^{\infty }A_{n}].  \label{11a}
\end{equation}

Matching both sides of Eq. (\ref{11a}) yields the following iterative
algorithm for the power series solution of Eq. (\ref{eqb}),
\begin{equation}
\mathcal{L}\left[ y_{0}\right] =\frac{as}{s^{2}+\omega ^{2}}-\frac{b^{2}}{%
s\left( s^{2}+\omega ^{2}\right) },  \label{12a}
\end{equation}%
\begin{equation}
\mathcal{L}\left[ y_{1}\right] =-\frac{1}{s^{2}+\omega ^{2}}\mathcal{L}\left[
A_{0}\right] ,  \label{12b}
\end{equation}%
\begin{equation}
\mathcal{L}\left[ y_{2}\right] =-\frac{1}{s^{2}+\omega ^{2}}\mathcal{L}\left[
A_{1}\right] ,  \label{12c}
\end{equation}%
\begin{equation*}
...
\end{equation*}%
\begin{equation}
\mathcal{L}\left[ y_{k+1}\right] =-\frac{1}{s^{2}+\omega ^{2}}\mathcal{L}%
\left[ A_{k}\right] .  \label{12n}
\end{equation}

By applying the inverse Laplace transformation to Eq. (\ref{12a}), we obtain
the value of $y_{0}$. Substituting $y_{0}$ into Eq. (\ref{Ad0}) to find the
first Adomian polynomial $A_{0}$. Then we substitute $A_{0}$ into Eq. (\ref%
{12b}), and we evaluate the Laplace transform of the quantities on the
right-hand side of it. The application of the inverse Laplace transformation
yields then the value of $y_{1}$. The other terms $y_{2}$, $y_{3}$, . . .,$%
y_{k+1}$, can be computed recursively in a similar step by step approach.

\subsection{The particular case $g(y)=\sum_{l=0}^{m}a_{l+2}y^{l+2}$}

Let us consider a nonlinear differential equation of the form
\begin{equation}
\frac{d^{2}y}{dx^{2}}+\omega ^{2}y+b^{2}+\sum_{l=0}^{m}a_{l+2}y^{l+2}=0,
\label{s1}
\end{equation}
where $\omega $, $b$,and $a_{l+2}$, $l=0,...,m$ are constants. Eq.~(\ref{s1}%
) must be integrated with the initial conditions $y(0)=y_{0}=a$, and $%
y^{\prime }(0)=0$, respectively. By applying the Laplace-Adomian method we
first take the Laplace transform to Eq.~(\ref{s1}), thus obtaining
\begin{equation}
\mathcal{L}\left( \frac{d^{2}y}{dx^{2}}\right) +\omega ^{2}\mathcal{L}\left(
y\right) +b^{2}\mathcal{L}\left( 1\right) +\sum_{l=0}^{m}a_{l+2}\mathcal{L}%
\left[ y^{l+2}\right] =0.
\end{equation}

Now we easily find
\begin{equation}
\mathcal{L}\left( y\right) \left( s^{2}+\omega ^{2}\right) =sy\left(
0\right) +y^{\prime }\left( 0\right) -\frac{b^{2}}{s}-\sum_{l=0}^{m}a_{l+2}%
\mathcal{L}\left[ y^{l+2}\right] =0,
\end{equation}
thus obtaining
\begin{equation}
\mathcal{L}\left( y\right) =\frac{sy\left( 0\right) +y^{\prime }\left(
0\right) }{s^{2}+\omega ^{2}}-\frac{b^{2}}{s\left( s^{2}+\omega ^{2}\right) }%
-\frac{1}{s^{2}+\omega ^{2}}\sum_{l=0}^{m}a_{l+2}\mathcal{L}\left[ y^{l+2}%
\right] .  \label{s2}
\end{equation}

From Eq.~(\ref{s2}) we obtain $y(x)$ in the form%
\begin{equation}
y\left( x\right) =\mathcal{L}^{-1}\left[ \frac{sy\left( 0\right) +y^{\prime
}\left( 0\right) }{s^{2}+\omega ^{2}}-\frac{b^{2}}{s\left( s^{2}+\omega
^{2}\right) }\right] -\mathcal{L}^{-1}\left[ \frac{1}{s^{2}+\omega ^{2}}%
\sum_{l=0}^{m}a_{l+2}\mathcal{L}\left[ y^{l+2}\right] \right] .
\end{equation}

We assume a power series solution for $y(x)$ as $y(x)=\sum_{n=0}^{\infty
}y_{n}(x)$, and we write the nonlinear terms as
\begin{equation}
y^{l+2}=\sum_{n=0}^{\infty }{A_{n,l+2}(x)},
\end{equation}%
where $A_{n,l+2}$ are the Adomian polynomial corresponding to $y^{l+2}$.
Then we obtain
\begin{equation}
\sum_{n=0}^{\infty }y_{n}(x)=\mathcal{L}^{-1}\left[ \frac{sy\left( 0\right)
+y^{\prime }\left( 0\right) }{s^{2}+\omega ^{2}}-\frac{b^{2}}{s\left(
s^{2}+\omega ^{2}\right) }\right] -\mathcal{L}^{-1}\left\{ \frac{1}{%
s^{2}+\omega ^{2}}\sum_{l=0}^{m}a_{l+2}\mathcal{L}\left[ \sum_{n=0}^{\infty }%
{A_{n,l+2}(x)}\right] \right\} .  \label{s3}
\end{equation}

We rewrite Eq.~(\ref{s3}) in the form
\begin{eqnarray}
y_{0}\left( x\right) +\sum_{n=0}^{\infty }y_{n+1}(x) &=&\mathcal{L}^{-1}%
\left[ \frac{sy\left( 0\right) +y^{\prime }\left( 0\right) }{s^{2}+\omega
^{2}}-\frac{b^{2}}{s\left( s^{2}+\omega ^{2}\right) }\right] -  \notag \\
&&\sum_{n=0}^{\infty }\mathcal{L}^{-1}\left\{ \frac{1}{s^{2}+\omega ^{2}}%
\sum_{l=0}^{m}a_{l+2}\mathcal{L}\left[ {A_{n,l+2}(x)}\right] \right\} .
\label{s4}
\end{eqnarray}

Eq.~(\ref{s4}) can be written as the recursive relations
\begin{equation}
y_{0}\left( x\right) =\mathcal{L}^{-1}\left[ \frac{sy\left( 0\right)
+y^{\prime }\left( 0\right) }{s^{2}+\omega ^{2}}-\frac{b^{2}}{s\left(
s^{2}+\omega ^{2}\right) }\right] ,
\end{equation}%
\begin{equation*}
...,
\end{equation*}

\begin{equation}
y_{k+1}(x)=-\mathcal{L}^{-1}\left\{ \frac{1}{s^{2}+\omega ^{2}}%
\sum_{l=0}^{m}a_{l+2}\mathcal{L}\left[ {A_{k,l+2}(x)}\right] \right\} .
\end{equation}

For the function $y^{l+2}$ a few Adomian polynomials are
\begin{equation}
A_{0,l+2}=y_{0}^{l+2},
\end{equation}%
\begin{equation}
A_{1,l+2}=(l+2)y_{1}y_{0}^{l+1},  \label{h2}
\end{equation}
\begin{equation}
A_{2,l+2}=(l+2)y_{2}y_{0}^{l+1}+(l+1)\left( l+2\right) \frac{y_{1}^{2}}{2!}%
y_{0}^{l},  \label{h3}
\end{equation}%
\begin{equation}
A_{3,l+2}=(l+2)y_{3}y_{0}^{l+1}+(l+1)\left( l+2\right)
y_{1}y_{2}y_{0}^{l}+l(l+1)\left( l+2\right) \frac{y_{1}^{3}}{3!}y_{0}^{l-1}.
\label{h4}
\end{equation}

For $k=0$ we obtain the first order approximation to the solution as
\begin{eqnarray}
y_{1}(x) &=&-\mathcal{L}^{-1}\left\{ \frac{\mathcal{L}\left[
a_{2}A_{0,2}(x)+a_{3}A_{0,3}(x)+a_{4}A_{0,4}(x)+...\right] }{s^{2}+\omega
^{2}}\right\} =  \notag \\
&&-\mathcal{L}^{-1}\left\{ \frac{\mathcal{L}\left(
a_{2}y_{0}^{2}+a_{3}y_{0}^{3}+a_{4}y_{0}^{4}+...\right) }{s^{2}+\omega ^{2}}%
\right\} .
\end{eqnarray}%
For $k=1$ we find%
\begin{eqnarray}
y_{2}(x) &=&-\mathcal{L}^{-1}\left\{ \frac{\mathcal{L}\left[
a_{2}A_{1,2}(x)+a_{3}A_{1,3}(x)+a_{4}A_{1,4}(x)+...\right] }{s^{2}+\omega
^{2}}\right\} =  \notag \\
&&-\mathcal{L}^{-1}\left\{ \frac{\mathcal{L}\left(
2a_{2}y_{1}y_{0}+3a_{3}y_{1}y_{0}^{2}+4a_{4}y_{1}y_{0}^{3}+...\right) }{%
s^{2}+\omega ^{2}}\right\} .
\end{eqnarray}%
$k=2$ gives%
\begin{eqnarray}
y_{3}(x) &=&-\mathcal{L}^{-1}\left\{ \frac{\mathcal{L}\left[
a_{2}A_{2,2}(x)+a_{3}A_{2,3}(x)+a_{4}A_{2,4}(x)+...\right] }{s^{2}+\omega
^{2}}\right\} =  \notag \\
&&-\mathcal{L}^{-1}\left\{ \frac{\mathcal{L}\left[ a_{2}\left(
2y_{2}y_{0}+y_{1}^{2}\right) +3a_{3}\left(
y_{2}y_{0}^{2}+y_{1}^{2}y_{0}\right) +a_{4}\left(
4y_{2}y_{0}^{3}+6y_{1}^{2}y_{0}^{2}\right) +...\right] }{s^{2}+\omega ^{2}}%
\right\} .
\end{eqnarray}%
Finally, for $k=3$ we obtain%
\begin{eqnarray}
y_{4}(x) &=&-\mathcal{L}^{-1}\Bigg\{\frac{\mathcal{L}\left[
a_{2}A_{3,2}(x)+a_{3}A_{3,3}(x)+a_{4}A_{3,4}(x)+...\right] }{s^{2}+\omega
^{2}}\Bigg\}=  \notag \\
&&-\mathcal{L}^{-1}\Bigg\{\frac{1}{s^{2}+\omega ^{2}}\mathcal{L}\Bigg[%
2a_{2}\left( y_{3}y_{0}+y_{1}y_{2}\right) +a_{3}\left(
3y_{3}y_{0}^{2}+6y_{1}y_{2}y_{0}+y_{1}^{3}\right) +  \notag \\
&&a_{4}\left( 4y_{3}y_{0}^{3}+12y_{1}y_{2}y_{0}^{2}+4y_{1}^{3}y_{0}\right)
+...\Bigg]\Bigg\}.
\end{eqnarray}%
Hence we have obtained the truncated power series solution of Eq.~(\ref{s1})
as given by%
\begin{equation}
y\left( x\right) =\sum_{n=0}^{\infty }y_{n}(x)=y_{0}\left( x\right)
+y_{1}\left( x\right) +y_{2}\left( x\right) +y_{3}\left( x\right)
+y_{4}\left( x\right) +....
\end{equation}

\section{The solution of the equation of the motion of the massive test
particles in Schwarzschild geometry by the Laplace Adomian Decomposition
Method}

\label{sect3}

In the following we will use a system of units with $G=c=1$. Then Eq. (\ref%
{eqSc}), describing the motion of a massive test particle in the
Schwarzschild geometry, takes the form
\begin{equation}
\frac{d^{2}u}{d\varphi ^{2}}+u=\frac{M}{L^{2}}+3Mu^{2}.  \label{a}
\end{equation}%
In order to simplify the mathematical formalism we rescale the function $u$
so that
\begin{equation}
u=\frac{1}{3M}U.
\end{equation}%
Thus Eq.~(\ref{a}) becomes
\begin{equation}  \label{a1}
\frac{d^{2}U}{d\varphi ^{2}}+U=b^{2}+U^{2},
\end{equation}%
where we have denoted $b^{2}=3M^{2}/L^{2}$. Eq.~(\ref{a1}) must be solved
with the initial conditions $U(0)=3Mu(0)=a$, and $U^{\prime }(0)=0$,
respectively.

\subsection{Power series solution of the equation of motion}

Assume that the solution of Eq. (\ref{a1}) can be obtained in power series
form,
\begin{equation}  \label{62}
U\left( \varphi \right) =\sum_{n=0}^{\infty }U_{n}\left( \varphi \right) .
\end{equation}

Now taking Laplace transform to Eq. (\ref{a1}) yields
\begin{equation}
\mathcal{L}\left[ \frac{d^{2}U}{d\varphi ^{2}}\right] +\mathcal{L}\left[ U%
\right] =b^2\mathcal{L}\left[ 1\right] +\mathcal{L}\left[ U^{2}\right] .
\end{equation}

Hence we obtain
\begin{equation}
s^{2}\mathcal{L}\left( U\right) -sU\left( 0\right) -U^{\prime }\left(
0\right) +\mathcal{L}\left( U\right) =\frac{b^{2}}{s}+\mathcal{L}\left[ U^{2}%
\right] ,
\end{equation}%
\begin{equation}
\mathcal{L}\left( U\right) =\frac{sU\left( 0\right) +U^{\prime }\left(
0\right) }{s^{2}+1}+\frac{b^{2}}{s\left( s^{2}+1\right) }+\frac{1}{s^{2}+1}%
\mathcal{L}\left[ U^{2}\right] .  \label{a2l}
\end{equation}

We write down a few Adomian polynomials for $U^{2}$,
\begin{equation}
A_{0}=U_{0}^{2},
\end{equation}%
\begin{equation}
A_{1}=2U_{1}U_{0},
\end{equation}%
\begin{equation}
A_{2}=2U_{2}U_{0}+U_{1}^{2},
\end{equation}%
\begin{equation}
A_{3}=2U_{3}U_{0}+2U_{1}U_{2},
\end{equation}%
Substituting Eq. (\ref{62}) and $U^{2}=\sum_{n=0}^{\infty }A_{n}\left(
\varphi \right) $ into Eq. (\ref{a2l}) gives the relation

\begin{equation}
\mathcal{L}\left[ \sum_{n=0}^{\infty }U_{n}\left( \varphi \right) \right] =%
\frac{sU\left( 0\right) +U^{\prime }\left( 0\right) }{s^{2}+1}+\frac{b^{2}}{%
s\left( s^{2}+1\right) }+\frac{1}{s^{2}+1}\mathcal{L}\left[
\sum_{n=0}^{\infty }A_{n}\left( \varphi \right) \right] ,
\end{equation}%
or, equivalently,

\begin{eqnarray}  \label{a2}
U_{0}\left( \varphi \right) +\sum_{n=1}^{\infty }U_{n}\left( \varphi \right)
&=&U_{0}\left( \varphi \right) +\sum_{n=0}^{\infty }U_{n+1}\left( \varphi
\right) =\mathcal{L}^{-1}\left[ \frac{sU\left( 0\right) +U^{\prime }\left(
0\right) }{s^{2}+1}+\frac{b^{2}}{s\left( s^{2}+1\right) }\right] +  \notag \\
&&\sum_{n=0}^{\infty }\mathcal{L}^{-1}\left[ \frac{\mathcal{L}\left[
A_{n}\left( \varphi \right) \right] }{s^{2}+1}\right] .
\end{eqnarray}

Next we rewrite Eq. (\ref{a2}) in the recursive forms
\begin{equation}
U_{0}\left( \varphi \right) =\mathcal{L}^{-1}\left[ \frac{sU\left( 0\right)
+U^{\prime }\left( 0\right) }{s^{2}+1}+\frac{b^{2}}{s\left( s^{2}+1\right) }%
\right] ,
\end{equation}%
\begin{equation*}
...,
\end{equation*}
\begin{equation}
U_{k+1}\left( \varphi \right) =\mathcal{L}^{-1}\left[ \frac{\mathcal{L}\left[
A_{k}\left( \varphi \right) \right] }{s^{2}+1}\right] .
\end{equation}

With the help of the explicit expressions of the Adomian polynomials, we
obtain
\begin{equation}
U_{0}(\varphi )=\left( a-b^{2}\right) \cos \varphi +b^{2},
\end{equation}%
\begin{equation}
A_{0}=U_{0}^{2}=\left[ \left( a-b^{2}\right) \cos \varphi +b^{2}\right] ^{2},
\end{equation}%
\begin{equation}
U_{1}\left( \varphi \right) =\mathcal{L}^{-1}\left[ \frac{\mathcal{L}\left[
A_{0}\left( \varphi \right) \right] }{s^{2}+1}\right] =\mathcal{L}^{-1}\left[
\frac{\mathcal{L}\left[ U_{0}^{2}\right] }{s^{2}+1}\right] ,k=0,
\end{equation}%
\begin{eqnarray}
U_{1}(\varphi ) &=&\frac{1}{6}\Bigg\{-2\left( a^{2}-2ab^{2}+4b^{4}\right)
\cos \varphi +3\left( a^{2}-2ab^{2}+3b^{4}\right) +  \notag \\
&&\left( a-b^{2}\right) \left[ \left( b^{2}-a\right) \cos (2\varphi
)+6b^{2}\varphi \sin \varphi \right] \Bigg\},
\end{eqnarray}%
\begin{eqnarray}
A_{1} &=&2U_{1}U_{0}=\frac{1}{3}\Bigg[\left( a-b^{2}\right) \cos \varphi
+b^{2}\Bigg]\Bigg\{-2\left( a^{2}-2ab^{2}+4b^{4}\right) \cos \varphi +
\notag \\
&&3\left( a^{2}-2ab^{2}+3b^{4}\right) +\left( a-b^{2}\right) \left[ \left(
b^{2}-a\right) \cos (2\varphi )+6b^{2}\varphi \sin \varphi \right] \Bigg\},
\end{eqnarray}%
\begin{equation}
U_{2}\left( \varphi \right) =\mathcal{L}^{-1}\left[ \frac{\mathcal{L}\left[
A_{1}\left( \varphi \right) \right] }{s^{2}+1}\right] =2\mathcal{L}^{-1}%
\left[ \frac{\mathcal{L}\left[ U_{1}U_{0}\right] }{s^{2}+1}\right] ,k=1,
\end{equation}%
\begin{eqnarray}
U_{2}\left( \varphi \right) &=&\frac{1}{144}\Bigg\{16\left(
a^{2}-5ab^{2}+7b^{4}\right) \left( a-b^{2}\right) \cos (2\varphi )+\cos
\varphi \Bigg[29a^{3}-183a^{2}b^{2}+  \notag \\
&&3ab^{4}\left( 125-24\varphi ^{2}\right) +b^{6}\left( 72\varphi
^{2}-509\right) \Bigg]+12\varphi \left(
5a^{3}-19a^{2}b^{2}+41ab^{4}-39b^{6}\right) \times  \notag \\
&&\sin \varphi -48\left( a^{3}-6a^{2}b^{2}+12ab^{4}-13b^{6}\right)
-48\varphi \left( b^{3}-ab\right) ^{2}\sin (2\varphi )+  \notag \\
&&3\left( a-b^{2}\right) ^{3}\cos (3\varphi )\Bigg\},
\end{eqnarray}%
\begin{equation}
U_{3}\left( \varphi \right) =\mathcal{L}^{-1}\left[ \frac{\mathcal{L}\left[
A_{2}\left( \varphi \right) \right] }{s^{2}+1}\right] =\mathcal{L}^{-1}\left[
\frac{\mathcal{L}\left[ 2U_{2}U_{0}+U_{1}^{2}\right] }{s^{2}+1}\right] ,k=2,
\end{equation}%
\begin{eqnarray}
U_{3}\left( \varphi \right) &=&\frac{1}{432}\Bigg\{-12\varphi \left(
5a^{3}-23a^{2}b^{2}+73ab^{4}-79b^{6}\right) \left( a-b^{2}\right) \sin
(2\varphi )-3\varphi \sin \varphi \Bigg[60a^{4}-  \notag \\
&&449a^{3}b^{2}+1335a^{2}b^{4}+3ab^{6}\left( 8\varphi ^{2}-721\right)
+b^{8}\left( 1721-24\varphi ^{2}\right) \Bigg]+\cos \varphi \times  \notag \\
&&\Bigg[-119a^{4}+4a^{3}b^{2}\left( 206-45\varphi ^{2}\right)
+3a^{2}b^{4}\left( 204\varphi ^{2}-961\right) +2ab^{6}\left( 2461-666\varphi
^{2}\right) +  \notag \\
&&4b^{8}\left( 279\varphi ^{2}-1262\right) \Bigg]-24\cos (2\varphi )\Bigg[%
4a^{4}-24a^{3}b^{2}+72a^{2}b^{4}-112ab^{6}-  \notag \\
&&6b^{4}\varphi ^{2}\left( a-b^{2}\right) ^{2}+63b^{8}\Bigg]+9\left(
25a^{4}-164a^{3}b^{2}+534a^{2}b^{4}-868ab^{6}+737b^{8}\right) +  \notag \\
&&27b^{2}\varphi \left( a-b^{2}\right) ^{3}\sin (3\varphi )+\left(
a-b^{2}\right) ^{4}(-\cos (4\varphi ))-  \notag \\
&&9\left( a-4b^{2}\right) \left( a-2b^{2}\right) \left( a-b^{2}\right)
^{2}\cos (3\varphi )\Bigg\},
\end{eqnarray}%
\begin{equation}
U_{4}\left( \varphi \right) =\mathcal{L}^{-1}\left[ \frac{\mathcal{L}\left[
A_{3}\left( \varphi \right) \right] }{s^{2}+1}\right] =2\mathcal{L}^{-1}%
\left[ \frac{\mathcal{L}\left[ U_{3}U_{0}+U_{1}U_{2}\right] }{s^{2}+1}\right]
,k=3,
\end{equation}%
\begin{eqnarray}
&&U_{4}(\varphi ) =\frac{1}{20736}\Bigg\{32\left(
2a^{2}-13ab^{2}+17b^{4}\right) \left( a-b^{2}\right) ^{3}\cos (4\varphi
)+108\varphi \Bigg(5a^{3}-27a^{2}b^{2}+105ab^{4}-  \notag \\
&&119b^{6}\Bigg)\left( a-b^{2}\right) ^{2}\sin (3\varphi )+36\left(
a-b^{2}\right) \cos (3\varphi )\Bigg(31a^{4}-196a^{3}b^{2}+642a^{2}b^{4}-
\notag \\
&&1036ab^{6}-54b^{4}\varphi ^{2}\left( a-b^{2}\right) ^{2}+595b^{8}\Bigg)%
+24\varphi \sin \varphi \Bigg[580a^{5}-4938a^{4}b^{2}+  \notag \\
&&a^{3}b^{4}\left( 20129-180\varphi ^{2}\right) +7a^{2}b^{6}\left( 84\varphi
^{2}-6583\right) +3ab^{8}\left( 20397-428\varphi ^{2}\right) +  \notag \\
&&5b^{10}\left( 204\varphi ^{2}-8221\right) \Bigg]+192\varphi \sin (2\varphi
)\Bigg[20a^{5}-223a^{4}b^{2}+884a^{3}b^{4}-2036a^{2}b^{6}+  \notag \\
&&2608ab^{8}+24b^{6}\varphi ^{2}\left( a-b^{2}\right) ^{2}-1289b^{10}\Bigg]%
+384\cos (2\varphi )\Bigg[16a^{5}+  \notag \\
&&a^{4}b^{2}\left( 15\varphi ^{2}-152\right) +8a^{3}b^{4}\left( 74-9\varphi
^{2}\right) +8a^{2}b^{6}\left( 27\varphi ^{2}-164\right) +104ab^{8}\left(
16-3\varphi ^{2}\right) +  \notag \\
&&b^{10}\left( 153\varphi ^{2}-862\right) \Bigg]-3\cos (\varphi )\Bigg[%
a^{5}\left( 600\varphi ^{2}-2357\right) +a^{4}b^{2}\left( 23209-5880\varphi
^{2}\right) +  \notag \\
&&2a^{3}b^{4}\left( 15588\varphi ^{2}-49577\right) +2a^{2}b^{6}\left(
126457-39324\varphi ^{2}\right) +ab^{8}\left( 96\varphi ^{2}\left(
1235-3\varphi ^{2}\right) -362713\right) +  \notag \\
&&b^{10}\left( 288\varphi ^{4}-83088\varphi ^{2}+298693\right) \Bigg]-288%
\Bigg( 50a^{5}-475a^{4}b^{2}+1934a^{3}b^{4}-4604a^{2}b^{6}+  \notag \\
&&6208ab^{8}-4337b^{10}\Bigg) - 192b^{2}\varphi \left( a-b^{2}\right)
^{4}\sin (4\varphi )+5\left( a-b^{2}\right) ^{5}\cos (5\varphi )\Bigg\}.
\end{eqnarray}

The power series solution of the equation of motion of the massive test
particles in Schwarzschild geometry is thus given by%
\begin{equation}  \label{77}
U\left( \varphi \right) =U_{0}\left( \varphi \right) +U_{1}\left( \varphi
\right) +U_{2}\left( \varphi \right) +U_{3}\left( \varphi \right)
+U_{4}\left( \varphi \right) +....,
\end{equation}%

\subsection{Comparison with the exact numerical solution}

\begin{figure}[!htb]
\centering
\includegraphics[scale=0.65]{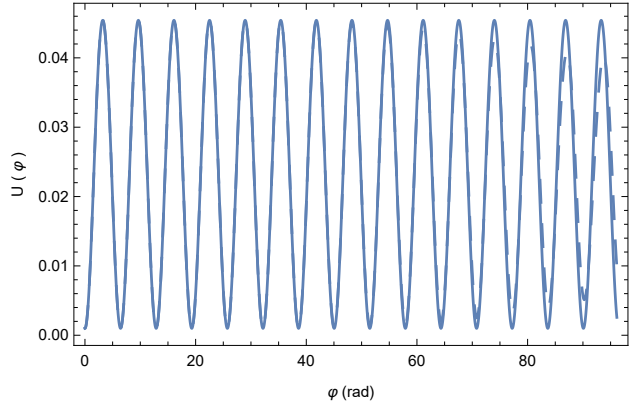}
\caption{Comparison of the numerical solution $U(\protect\varphi)$ of Eq.~(%
\protect\ref{a1}) (solid curve), and of the truncated power series solution
obtained by the Laplace Adomian Decomposition method, given by Eq.~(\protect
\ref{77}) (dashed curve), for $b^2=0.0225$ and $a=0.001$.}
\label{fig1}
\end{figure}

In order to estimate the results obtained by the Laplace Adomian
Decomposition Method, in Fig.~\ref{fig1} we present the comparison between
the exact numerical solution of Eq.~(\ref{a1}), and the analytical, power
series representation, given by Eq.~(\ref{77}), for $b=0.15$, $b^2=0.0225$,
and $a=0.001$, respectively.

\begin{figure}[!htb]
\centering
\includegraphics[scale=0.65]{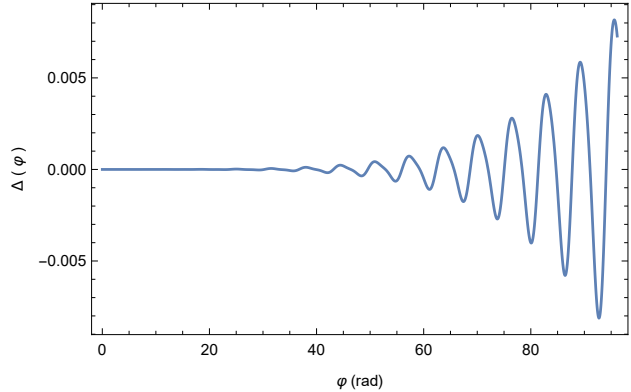}
\caption{The absolute difference $\Delta (\protect\varphi)$ between the
numerical solution $\left.U(\protect\varphi)\right|_{\mathrm{num}}$ and the
truncated Laplace-Adomian power series solution $\left.U(\protect\varphi%
)\right|_{\mathrm{LADM}}$ of Eq.~(\protect\ref{a1}) for $b^2=0.0225$ and $%
a=0.001$.}
\label{fig2}
\end{figure}

The absolute difference $\Delta (\varphi)$, defined as
\begin{equation}
\Delta (\varphi)=\left.U(\varphi)\right|_{\mathrm{num}}-\left.U(\varphi)%
\right|_{\mathrm{LADM}},
\end{equation}
between the numerical solution $\left.U(\varphi)\right|_{\mathrm{num}}$ and
the truncated Laplace-Adomian power series solution is represented in Fig.~%
\ref{fig2}.

\subsection{Application: the motion of planet Mercury}

In order to integrate the equation of motion of the massive test particles
in Schwarzschild geometry we need to know the initial conditions of the
motion. For the initial value of $u$ and $U$ we adopt the values
\begin{equation}
u(0)=\frac{M}{L^{2}}\left( 1+e\right) ,a=U(0)=\frac{3M^{2}}{L^{2}}\left(
1+e\right) =b^{2}\left( 1+e\right) ,
\end{equation}%
where $e$ is the eccentricity of the orbit. The angular momentum $L^{2}$ can
be expressed in terms of the geometric parameters of the orbit by using Eq.~(%
\ref{L2}). Thus in physical units we obtain
\begin{equation}
b^{2}=\frac{3GM}{c^{2}\bar{a}\left( 1-e^{2}\right) },a=\frac{3GM}{c^{2}\bar{a%
}\left( 1-e\right) }.
\end{equation}

Therefore for this choice of parameters and initial conditions the
successive terms in the power series solution of the equation of motion can
be obtained by the Laplace Adomian decomposition method as follows:
\begin{equation}
u_{0}\left( \varphi \right) =\frac{1+e\cos \varphi }{\bar{a}\left(
1-e^{2}\right) },
\end{equation}%
\begin{equation}
u_{1}\left( \varphi \right) =-\frac{GM\left[ -3\left( e^{2}+2e\varphi \sin
\varphi +2\right) +e^{2}\cos (2\varphi )+2\left( e^{2}+3\right) \cos \varphi %
\right] }{2c^{2}\bar{a}^{2}\left( 1-e^{2}\right) ^{2}},
\end{equation}%
\begin{eqnarray}
u_{2}\left( \varphi \right) &=&\frac{G^{2}M^{2}}{16c^{4}\bar{a}^{3}\left(
e^{2}-1\right) ^{3}}\Bigg\{e\Bigg[-3e^{2}\cos (3\varphi )-16((e-3)e+3)\cos
(2\varphi )+  \notag \\
&&12\varphi \sin \varphi (8e\cos \varphi +(4-5e)e-18)\Bigg]+\Bigg[e\left(
(96-29e)e+72\varphi ^{2}-96\right) +288\Bigg]\times  \notag \\
&&\cos \varphi +48(e((e-3)e+3)+3\varphi \sin \varphi -6)\Bigg\},
\end{eqnarray}%
\begin{eqnarray}
u_{3}\left( \varphi \right) &=&\frac{G^{3}M^{3}}{16c^{6}\bar{a}^{4}\left(
1-e^{2}\right) ^{4}}\Bigg\{e^{4}\left( -\cos (4\varphi )\right)
+27e^{3}\varphi \sin (3\varphi )-24\Bigg[-6e^{2}\varphi ^{2}+  \notag \\
&&4e(e((e-2)e+6)-6)+3\Bigg]\cos (2\varphi )-9(e-3)(e-1)e^{2}\cos (3\varphi )+
\notag \\
&&\Bigg[-119e^{4}+348e^{3}-1125e^{2}-36(e(e(5e-2)+18)-6)\varphi
^{2}+1152e-2304\Bigg]\cos \varphi -  \notag \\
&&3\varphi \Bigg[e\left( e(e(60e-209)+348)+24\left( \varphi ^{2}-25\right)
\right) +504\Bigg]\sin \varphi +  \notag \\
&&12\left[ e((8-5e)e-42)+24\right] e\varphi \sin (2\varphi )+9\left[
e(e(e(25e-64)+192)-192)+264\right] \Bigg\},  \notag \\
&&
\end{eqnarray}%
\begin{eqnarray}
u_{4}\left( \varphi \right) &=&\frac{G^{4}M^{4}}{256c^{8}\bar{a}^{5}\left(
e^{2}-1\right) ^{5}}\Bigg\{-5e^{5}\cos (5\varphi )+192e^{4}\varphi \sin
(4\varphi )-32(e(2e-9)+6)\times  \notag \\
&&e^{3}\cos (4\varphi )-108\left[ e(e(5e-12)+66)-36\right] e^{2}\varphi \sin
(3\varphi )+24\varphi \Bigg[-580e^{5}+2038e^{4}-  \notag \\
&&6177e^{3}+9522e^{2}+12(e(e(15e-4)+54)-12)\varphi ^{2}-12564e+10224\Bigg]%
\sin (\varphi )+  \notag \\
&&3\Bigg[-2357e^{5}+11424e^{4}-29888e^{3}+71136e^{2}-288e\varphi ^{4}+
\notag \\
&&24(e(e(e(5e(5e-24)+569)-600)+1428)-720)\varphi ^{2}-73296e+110592\Bigg]%
\cos (\varphi )-  \notag \\
&&192\varphi \Bigg[e\left( e\left( e(e(20e-123)+192)+6\left( 4\varphi
^{2}-87\right) \right) +396\right) -36\Bigg]\sin (2\varphi )-  \notag \\
&&36e\Bigg[e\left( e\left( e(31e-72)-54\varphi ^{2}+240\right) -216\right)
+36\Bigg]\cos (3\varphi )-  \notag \\
&&384\Bigg[3e(e(e(5e-4)+30)-12)\varphi ^{2}+8e(e(e(e(2e-9)+18)-36)+36)-54%
\Bigg]\cos (2\varphi )+  \notag \\
&&288\Bigg[e(e(e(25e(2e-9)+534)-1152)+1152)-1224\Bigg]\Bigg\}.
\end{eqnarray}

In the case of the planet Mercury its orbital parameters are $\bar{a}%
=57.91\times 10^{11}$ cm, and $e=0.205615$, respectively. $M$ is the mass of
the Sun, given by $M=M_{\odot}=1.989\times 10^{33}$ g. Therefore the
successive approximations of the solution of the equation of motion of
planet Mercury in the Schwarzschild metric created by the gravitational
field of the Sun can be obtained as
\begin{equation}
u_0(\varphi)=1.80305\times 10^{-13} \left(1+0.205615 \cos \varphi \right),
\end{equation}
\begin{eqnarray}
u_1\left(\varphi \right)&=&-2.39927\times{10}^{-21} \Bigg[-3 (0.41123
\varphi \sin \varphi +2.04228)+6.08456 \cos \varphi +  \notag \\
&& 0.0422775 \cos (2 \varphi )\Bigg],
\end{eqnarray}
\begin{eqnarray}
u_2\left(\varphi \right)&=&-7.98169\times 10^{-30} \Bigg\{\Bigg[0.205615
\left(72 \varphi ^2-77.487\right)+288.00\Bigg] \cos (\varphi )+  \notag \\
&& 48 \left(3 \varphi \sin \varphi -5.50129\right)+0.205615 \Bigg[-38.8069
\cos (2 \varphi )-0.126833 \cos (3 \varphi )+  \notag \\
&& 12 \varphi \sin \varphi (1.64492 \cos \varphi -17.3889)\Bigg]\Bigg\},
\end{eqnarray}
\begin{eqnarray}
u_3\left(\varphi\right)&=&2.12421\times 10^{-37} \Bigg\{-3 \varphi \Bigg[%
0.205615 \left(24 \left(\varphi ^2-25\right)+63.2396\right)+504\Bigg] \sin
\varphi +  \notag \\
&& \left(84.2407 \varphi ^2-2111.88\right) \cos \varphi -24 \left(-0.253665
\varphi ^2-0.982493\right) \cos (2 \varphi )+  \notag \\
&& 38.6365 \varphi \sin (2 \varphi )+0.234708 \varphi \sin (3 \varphi
)-0.844636 \cos (3 \varphi )-  \notag \\
&& 0.00178739 \cos (4 \varphi )+2089.15\Bigg\},
\end{eqnarray}
\begin{eqnarray}
u_4\left(\varphi\right)&=&-3.53332\times 10^{-46} \Bigg\{24 \varphi
\left(7992.95\, -11.2261 \varphi ^2\right) \sin \varphi -  \notag \\
&& 7.40214 \Bigg[0.205615 \left(0.205615 \left(226.506\, -54 \varphi
^2\right)-216\right)+36\Bigg] \cos (3 \varphi )-  \notag \\
&& 192 \varphi \Bigg[0.205615 \left(0.205615 \left(6 \left(4 \varphi
^2-87\right)+34.4518\right)+396\right)-36\Bigg] \sin (2 \varphi )-  \notag \\
&& 384 \left(-3.67467 \varphi ^2-5.82984\right) \cos (2 \varphi )+3
\left(-59.2171 \varphi ^4-10728.2 \varphi ^2+98288.4\right) \cos \varphi +
\notag \\
&& 104.53 \varphi \sin (3 \varphi )+0.343179 \varphi \sin (4 \varphi
)-1.17779 \cos (4 \varphi )-0.00183757 \cos (5 \varphi )-  \notag \\
&& 297094.\Bigg\}.
\end{eqnarray}
\begin{eqnarray}
u\left(\varphi\right)&=&\left(6.88343\times 10^{-44} \varphi
^3+8.20724\times 10^{-36} \varphi \right) \sin (2 \varphi )+\sin \varphi %
\Bigg(-3.14475\times {10}^{-36} \varphi ^3+  \notag \\
&& 2.95996\times 10^{-21} \varphi -3.23948\times 10^{-29} \varphi \cos
\varphi \Bigg)+\left(1.29321\times 10^{-36} \varphi ^2+5.00886\times
10^{-36}\right) \times  \notag \\
&& \cos (2 \varphi )+\left(-5.97095\times 10^{-45} \varphi ^2-1.79419 \times
10^{-37}\right) \cos (3 \varphi )+\Bigg(6.27699\times 10^{-44} \varphi ^4-
\notag \\
&& 1.18163 \times 10^{-28} \varphi ^2+3.70733\times 10^{-14}\Bigg) \cos
\varphi +4.98570\times 10^{-38} \varphi \sin (3 \varphi )-  \notag \\
&& 1.21256 \times 10^{-46} \varphi \sin (4 \varphi )-1.01435\times 10^{-22}
\cos (2 \varphi )+2.08152\times 10^{-31} \cos (3 \varphi )-  \notag \\
&& 3.79680\times 10^{-40} \cos (4 \varphi )+6.49272\times 10^{-49} \cos (5
\varphi )+1.80304\times 10^{-13}.
\end{eqnarray}

\subsubsection{The perihelion precession}

In the first order approximation the solution of the equation of motion for
massive particles in Schwarzschild geometry is obtained as
\begin{eqnarray}
u(\varphi ) &\approx &u_{0}(\varphi )+u_{1}(\varphi )=\frac{1+e\cos \varphi
}{\bar{a}\left( 1-e^{2}\right) }+  \notag \\
&&\frac{GM\left[ 3\left( e^{2}+2e\phi \sin \varphi +2\right) -e^{2}\cos
(2\varphi )-2\left( e^{2}+3\right) \cos \varphi \right] }{2\bar{a}%
^{2}c^{2}\left( 1-e^{2}\right) ^{2}}.
\end{eqnarray}

By neglecting the constant terms and those who oscillate through two cycles
on each orbit we obtain

\begin{equation}
u(\varphi )\approx \frac{1+e\cos \varphi +e\alpha \varphi \sin \varphi }{l}=%
\frac{1+e\cos \left[ \left( 1-\alpha \right) \varphi \right] }{l},
\end{equation}%
where we have denoted $l=\bar{a}\left( 1-e^{2}\right) $, and $\alpha
=3GM/c^{2}l<<1$. At perihelion we now have $\left( 1-\alpha \right) \varphi
=2n\pi $, where $n$ is an integer, or $\varphi =2n\pi +6n\pi \left(
GM/lc^{2}\right) $. This relation shows that the perihelion advances by $%
\Delta \varphi =6n\pi \left( GM/lc^{2}\right) $ per rotation, and the rate
of precession is
\begin{equation}
\frac{\Delta \varphi }{\Delta t}=\frac{6\pi GM}{\bar{a}c^{2}\left(
1-e^{2}\right) \Delta t},n=1.
\end{equation}

We have thus recovered from the first order Laplace-Adomian solution of the
equation of motion the standard general relativistic result for the
perihelion precession \cite{LaLi}.

\section{Solving the equation of motion for light in Schwarzschild geometry}

\label{sect4}

The equation of motion of light in Schwarzschild geometry is a particular
case of the general equation of motion (\ref{a1}) for $b^2=0$, and it is
given by
\begin{equation}  \label{defphot}
\frac{d^2U}{d\varphi ^2}+U=U^2.
\end{equation}

\subsection{Power series solution}

The truncated power series solution of Eq.~(\ref{defphot}) can be obtained
immediately by using the Laplace-Adomian decomposition method as
\begin{equation}
U_0(\varphi)=a \cos \varphi ,
\end{equation}
\begin{equation}
U_1(\varphi)=\frac{2}{3} a^2 \sin ^2\frac{\varphi }{2} \left(\cos \varphi
+2\right),
\end{equation}
\begin{equation}
U_2(\varphi)=\frac{1}{144} a^3 \left[60 \varphi \sin \varphi +29 \cos
\varphi +16 \cos (2 \varphi )+3 \cos (3 \varphi )-48\right],
\end{equation}
\begin{eqnarray}
U_3(\varphi)&=&-\frac{1}{432} a^4 \Bigg[180 \varphi \sin \varphi +60 \varphi
\sin (2 \varphi )+119 \cos \varphi +96 \cos (2 \varphi )+9 \cos (3 \varphi )+
\notag \\
&& \cos (4 \varphi )-225\Bigg],
\end{eqnarray}
\begin{eqnarray}
U_4\left(\varphi\right)&=&\frac{a^5}{20736} \Bigg\{\left(7071-1800 \varphi
^2\right) \cos \varphi +6144 \cos (2 \varphi )+1116 \cos (3 \varphi )+64
\cos (4 \varphi )+  \notag \\
&& 5 \Bigg[12 (232 \varphi \sin \varphi +64 \varphi \sin (2 \varphi )+9
\varphi \sin (3 \varphi )-240)+\cos (5 \varphi )\Bigg]\Bigg\},
\end{eqnarray}
\begin{eqnarray}  \label{solphot}
U(\varphi)&=&\frac{a}{20736} \Bigg\{a \Bigg[384 \left(2 a \left(8 a^2-6
a+3\right)-9\right) \cos (2 \varphi )+a (36 (a (31 a-12)+12)\times  \notag \\
&& \cos (3 \varphi )+a \Bigg(16 \left(4 a-3\right) \cos (4 \varphi )+5 a
\cos (5 \varphi )+60 \varphi \sin \varphi \Big(18 a \cos (2 \varphi )+
\notag \\
&& 32 (4 a-3) \cos \varphi +241 a-144\Big)\Bigg)+8640 \varphi \sin \varphi
)-144 \Bigg(a \Big(25 a (4 a-3)+48\Big)-72\Bigg)\Bigg]+  \notag \\
&& 3 \Bigg[a \left(a \left(a \left(a \left(2357-600 \varphi
^2\right)-1904\right)+1392\right)-2304\right)+6912\Bigg] \cos \varphi \Bigg\}%
.
\end{eqnarray}

\begin{figure}[!htb]
\centering
\includegraphics[scale=0.65]{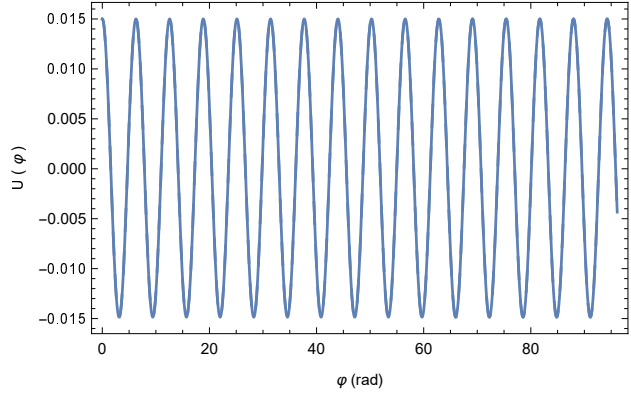}
\caption{Comparison of the numerical solution $U(\protect\varphi)$ of Eq.~(%
\protect\ref{defphot}), describing the motion of a photon in Schwarzschild
geometry (solid curve), and of its truncated power series solution obtained
by the Laplace Adomian Decomposition method, given by Eq.~(\protect\ref%
{solphot}) (dashed curve), for $a=0.015$.}
\label{fig3}
\end{figure}

\begin{figure}[!htb]
\centering
\includegraphics[scale=0.65]{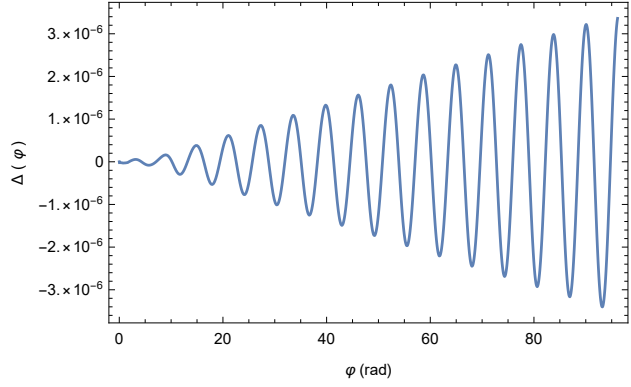}
\caption{The absolute difference $\Delta (\protect\varphi)$ between the
numerical solution $\left.U(\protect\varphi)\right|_{\mathrm{num}}$ and the
truncated Laplace-Adomian power series solution $\left.U(\protect\varphi%
)\right|_{\mathrm{LADM}}$ of Eq.~(\protect\ref{defphot}) for $a=0.015$.}
\label{fig4}
\end{figure}

The comparison of the numerical solution of Eq.~(\ref{defphot}) and of its
truncated power series solution (\ref{solphot}), as well as their
difference, are presented in Figs.~\ref{fig3} and \ref{fig4}, respectively.

\subsection{The bending angle of light}

In the Newtonian approximation the solution of the equation of motion for
the light has the solution $u_{0}=\cos \varphi /R$, where $R$ is the
distance of the closest approach to the massive object. This fixes the value
of the constant $a$ as $a=3GM/c^{2}R$. In the first order of approximation
the equation of motion of the photon in the Schwarzschild geometry has the
solution
\begin{equation}
u(\varphi )\approx u_{0}(\varphi )+u_{1}(\varphi )=\frac{1}{R}\cos \varphi +%
\frac{2GM}{c^{2}R^{2}}\sin ^{2}\frac{\varphi }{2}\left( 2+\cos \varphi
\right) .
\end{equation}%
By taking $\varphi =\pi /2+\epsilon $, we have
\begin{equation}
u(\varphi )\approx \frac{GM}{2c^2R^2}\left\{2 \left(1-\frac{c^2 R}{GM}\right)\sin \epsilon  + \left[\cos (2 \epsilon )+3\right]\right\}.
\end{equation}

By performing a first order series expansion with respect to $\epsilon $,
and taking $u=0$, gives the light deflection angle as
\begin{equation}
\delta =2\epsilon \approx \frac{4GM}{c^{2}R\left( 1-GM/c^{2}R\right) },
\end{equation}%
which in the limit $GM/c^{2}R<<1$ reduces to the well-known general
relativistic result $\delta \approx 4GM/c^{2}R$ \cite{LaLi}. In the second
order of approximation, with $u\approx u_{0}+u_{1}+u_{2}$, we obtain by
using the same procedure
\begin{equation}
\delta =\frac{4GM}{c^{2}R}\frac{1+\left( \frac{15\pi }{16}-2\right) \frac{GM%
}{c^{2}R}}{1-\frac{GM}{c^{2}R}-\frac{5G^{2}M^{2}}{2c^{4}R^{2}}}.
\end{equation}

Finally, in the fourth order of approximation we obtain
\begin{equation}
\delta \approx \frac{4GM}{c^{2}R}\frac{1+\frac{(15\pi -32)}{16}\frac{GM^{{}}%
}{c^{2}R}-\frac{5(9\pi -32)}{16}\frac{G^{2}M^{2}}{c^{4}R^{2}}+\frac{5(669\pi
-2048)}{256}\frac{G^{3}M^{3}}{c^{6}R^{3}}}{1-\frac{GM}{c^{2}R}-\frac{%
5G^{2}M^{2}}{2c^{4}R^{2}}-\frac{(15\pi -22)}{4}\frac{G^{3}M^{3}}{c^{6}R^{3}}-%
\frac{\left( 4816-1920\pi +225\pi ^{2}\right) }{128}\frac{G^{4}M^{4}}{%
c^{8}R^{4}}}.
\end{equation}

\section{Discussions and concluding remarks}

\label{sect5}

In the present paper we have considered the applications of a very powerful
mathematical method, the Laplace-Adomian Decomposition Method, for the study
of the general relativistic equations describing the motions of massive test
particles and of photons in spherical symmetric geometries. The
Laplace-Adomian Decomposition Method has been used extensively for obtaining
power series solutions of many classes of nonlinear differential and
integral equations. High precision solutions can be obtained by using only a
few terms in a truncated series expansion. The main "ingredients" in this
method are the Adomian polynomials, which are generated recursively in a
step by step procedure.

After introducing a general formalism for obtaining the equations of motion
in arbitrary spherically symmetric spacetimes, and after the presentation of
the Laplace-Adomian Decomposition Method, we have investigated in detail the
dynamics of massive particles and photons in Schwarzschild geometry. The
equations of motion can be easily solved, and we have obtained their series
solution. One of the advantages of the Laplace-Adomian Decomposition Method
is that it gives directly the solution in terms of trigonometric functions,
and not in terms of power series of the angular variable $\varphi $. A
truncated series solution containing only five terms can reprduce the exact
numerical solution with a high precision. Moreover, obtaining the terms in
the Adomian series can be easily done with the use of some symbolic
software. The basic known physical results can be easily reobtained in the
first order approximation, as well as for higher orders. For the perihelion
precession we have investigated only the first order effects, but in the
case of the deflection of light we have obtained the bending angle up to the
fourth order of approximation.

\bigskip Several important generalizations and extensions of the
Schwarzschild metric are known. In the case of the vacuum spacetimes outside
charged spherically symmetric objects carrying a charge $Q$ the geometry is
given by the Reissner-Nordstrom metric,
\begin{equation}
e^{\nu }=e^{-\lambda }=1-\frac{2GM}{c^2r}+\frac{Q^{2}}{r^{2}}.
\end{equation}

In the case of rotating star the first order monopole correction to the
Schwarzschild metric is given by \cite{Glen}
\begin{equation}
e^{\nu }=e^{-\lambda }=1-\frac{2GM}{c^2r}+\frac{J^{2}}{r^{4}},
\end{equation}%
where $J=I\Omega $, where $I$ is the moment of inertia, $\Omega $ is the
angular velocity, and $J$ is the angular momentum of the star, respectively.

The physics of the particle motion in these metrics can be easily
investigated by using the methods developed in the present paper. For
example, in the case of the Reissner-Nordstrom metric, $%
f(u)=2GMu/c^{2}-Qu^{2}$,
\begin{equation}
G(u)=2\frac{GM}{c^{2}}u^{3}-Qu^{4}+\frac{E^{2}}{c^{2}L^{2}}-\frac{1}{L^{2}}+%
\frac{2GMu}{c^{2}L^{2}}-\frac{Qu^{2}}{L^{2}},
\end{equation}%
and
\begin{equation}
F(u)=\frac{3GM}{c^{2}}u^{2}-2Qu^{3}+\frac{GM}{c^{2}L^{2}}-\frac{Qu}{L^{2}},
\end{equation}%
respectively. Therefore the equation of motion in the Reissner-Nordstrom
metric is given by
\begin{equation}
\frac{d^{2}u}{d\varphi ^{2}}+\left( 1+\frac{Q}{L^{2}}\right) u=\frac{3GM}{%
c^{2}}u^{2}-2Qu^{3}+\frac{GM}{c^{2}L^{2}},
\end{equation}%
which can be solved easily by using the Laplace-Adomian Decomposition
Method. The radius of the circular orbit $u_{0}$ can be obtained as a
solution of the algebraic equation
\begin{equation}
\frac{3GM}{c^{2}}u_{0}^{2}-u_{0}+\frac{GM}{c^{2}L^{2}}=2Qu_{0}^{3}+\frac{Q}{%
L^{2}}u_{0}.  \label{u0DMPR}
\end{equation}%
In the first order of approximation, and with the assumption $Q/L^{2}\ll 1$,
$u_{0}$ can be approximated as $u_{0}\approx GM/c^{2}L^{2}$. Thus, the
perihelion precession $\Delta \phi $ in the Reissner-Nordstrom geometry is
given by
\begin{equation}
\Delta \phi =\frac{6\pi GM}{c^{2}a(1-e^{2})}-\frac{\pi c^{2}Q}{GMa(1-e^{2})}.
\label{delsol}
\end{equation}%
The first term in Eq. (\ref{delsol}) gives the general relativistic
correction term for the perihelion precession, while the second term gives
the correction due to the presence of the charge. In the case of the motion
of massless particles we have
\begin{equation}
P(u)=2\frac{GM}{c^{2}}u^{3}-Qu^{4}+\frac{E^{2}}{c^{2}L^{2}},
\end{equation}%
and
\begin{equation}
V(u)=\frac{3GM}{c^{2}}u^{2}-2Qu^{3}.
\end{equation}%
Hence the equation of motion for photons is given by
\begin{equation}
\frac{d^{2}u}{d\phi ^{2}}+u=\frac{3GM}{c^{2}}u^{2}-2Qu^{3}.
\end{equation}%
This equation can also be investigated easily by using the Laplace Adomian
Decomposition Method, and the methods developed in the present paper.

\end{document}